\documentstyle[12pt]{article}
\def\beq{\begin{equation}}
\def\eeq{\end{equation}}
\def\ben{\begin{eqnarray}}
\def\een{\end{eqnarray}}
\def\bea{\begin{array}}
\def\eea{\end{array}}

\begin{document}

\baselineskip=20pt

%------------------------------------------------------------------------
%   BEGIN THE TITLEPAGE
%------------------------------------------------------------------------

\begin{flushright}
Prairie View A \& M, HEP-3-95\\
March 1995%\hskip

\end{flushright}

\vskip.15in

\begin{center}
{\Large\bf Mass hierarchy and the spectrum  of scalars
 }\\

\vskip .5in

{\bf Dan-di Wu}\footnote{ $~$E-mail: DANWU@
physics.rice.edu or wu@hp75.pvamu.edu}

{\sl HEP, Prairie View A\&M University, Prairie View, TX 77446-0355, USA}

\vskip .3in
Submitted to Physical Review Letters
\end{center}
\vskip.4in

{\bf Abstract:}\\

We use the natural $SU(3)\times U(1)$ global symmetry of the gauge-fermion
interaction sector  of the standard model to discuss the fermion mass
hierarchy problem. The $SU(3)$ sixtet and triplet Higgs are introduced.
  The Yukawa sector is partially symmetric. The smaller
the symmetry of a Yukawa term, the smaller its coupling constant. The mass
hierarchy is a combined effect of smaller coupling  constants and smaller
VEVs. There is a bunch of pseudo-goldstone bosons  which obtains their
masses
mainly from the small explicit breaking terms in the Higgs potential.

\newpage
The recent reports from CDF and D0 further confirm the existence of a
heavy
top quark[1]. Up to the present every particle in the minimal standard
model[2,3,4]
has been found except for the Higgs particle[5]. However many problems
are still
unresolved for the minimal standard model (MSM). Prominent among them
is why
quarks and leptons have  specific hierarchical  masses and  very small
mixing.
Another unresolved problem is  the exact mechanism of the electroweak
symmetry
breakdown, which, because of the lack of  evidence of a Higgs particle,
is the most uncertain in the MSM.

To attack the above two
problems by relating  one to another, S. Weinberg has proposed a multi-Higgs
doublet model[6]. Noting that the mass of the top is, within a factor of two,
close to that of the weak gauge bosons, and all the other quarks and leptons
are much lighter,
Weinberg assumes that the top ($t$) and the weak gauge bosons, $W$ and $Z$,
obtain
their masses mainly from the same  vacuum expectation value (VEV) of
a Higgs
multiplet. The specific multiplet he chooses is a $SU(2)_L$ doublet, a
nd a $SU(3)$
triplet at the same time. The $SU(3)$ here is a global symmetry of
his model.
The Yukawa coupling term in the Lagrangain which  serves the top a
mass is $SU(3) $
symmetric before a spontaneously symmetry breakdown (SSB). Therefore
its coupling
constant can be at the order of 1. The $b$ quark
also obtains its mass from the same expectation value. However the Yukawa
coupling term which serves its mass is originally $SO(3)$ symmetric, where
$SO(3)$ is a subgroup of the complete global symmetry $SU(3)$.
Its coupling constant is therefore smaller, because the coupling term
is less
symmetric. All the other quarks and leptons should get their masses
from other
sources which do not enjoy such large global symmetries, and are
from different
VEVs (which are smaller). Indeed it is more natural to assume that
masses with
different orders of magnitude  come from different sources than to
account for
them by one arbitrarily adjustable Yukawa coupling constant, which in
the minimal
standard model (which has the simplest possible scalar spectrum) runs from
10$^{-6}$ for the electron to 10$^{0}$ for the top. He then discusses the
properties of the pseudo-goldstone
bosons from the spontaneously symmetry breakdown of $SU(3)$. In
particular, he
points out that there is not a
$Z-Z-PGB$ coupling where $PGB$ represents a pseudo-goldstone boson.
Therefore,
for instance, the LEP experiment cannot put a mass limit on these
light particles,
 no matter how light they are,
 although their masses are at the order of 10$^2$ GeV, according
 to Weinberg.
Thus Weinberg explains, in some sense, why the top quark is much
heavier than the
other quarks and leptons
and the beauty is the next heaviest, and he predicts a copious Higgs
spectrum,
in particular, a bunch of $PGBs$.

In this note we will devote ourselves to a similar line of thinking. We will
introduce two modifications to Weinberg's original model:\\
1.  We take
\beq
{\bf G}=SU(3)\times U(1)
\eeq
 as the global symmetry of the main part of the Lagrangain. We think
 that this is more
natural, because the gauge-fermion interaction sector of the
Lagrangian enjoys
this bigger global symmetry. The basis of this symmetry lies in the
fact that
there are three families of quarks and leptons. In addition we  follow
Weinberg
to
assume that members of one family can transform differently under
the $SU(3)$
transformation. In other words,  the three left-handed doublets of
quarks are
in a $SU(3)$ triplet, while the three right-handed up (or down) type
quarks may
be, as in the model we are presenting, in a $SU(3)$ anti-triplet.
Therefore the
global symmetry $SU(3)\times U(1) $, which we are talking about here is
completely different from a family global symmetry. For a family group, all
members in one family are collectively one object of the group
transformation.
In the sense of its changing an object in one family to another,  this
global symmetry
 is a horizontal symmetry.\\
2. We have one triplet
Higgs as well as one sixtet Higgs. The sixtet  Higgs will develop a
large VEV,
while the triplet, a small VEV.
The advantage of introducing the sixtet Higgs is that its big VEV
naturally
contributes to the big mass of the top and the relatively big mass of the
beauty[6]. Further more,
 it explains  the smallness of
 the weak
mixing between the third and other generations  straight forwardly
(see later). The triplet Higgs in this model is instead responsible for
 the masses of
the c and s quarks. An approximate value of $V_{cb}$ is then calculable.

The bigger global symmetry ${\bf G} $ allows us to have more explicit
symmetry
breaking terms in the Yukawa sector.
We can write, in addition to an $SO(3)$ symmetric Yukawa coupling term,
 also $SU(3)$ and $SU(2) \times U(1)^\prime\times Z_2$ symmetric coupling
  terms. We therefore
will be able to reproduce a mass matrix which is close to the one
proposed by
Fritzsch[7]. Since the VEVs of both Higgs multiplets contribute to
masses of the quarks with the same electric charges, our model will have
flavor-changed neutral currents mediated by
pseudo-goldstone bosons at the tree level. However, as discussed by many
authors, it should not be very difficult to meet the most crucial
experimental
limits on flavor-changed neutral currents, if care is taken in model
building[10].

As we emphasized  the sector in our model which involves gauge interactions
is completely
standard, therefore, we will not discuss it. We will concentrate ourselves
on the Higgs potential sector and the Yukawa sector.
After presenting the model and exploring some of its features, we will
briefly
discuss the discovery channels for pseudo-goldstone bosons.

First let us give the fermion and scalar contents of the model in
{\bf Table 1},
where $i$ and $j$ are $SU(3)$ indices which run from 1 to 3.
All fields, except the
standard gauge fields, and their global and gauge quantum numbers
are listed in this table.  From {\bf
Table 1} we see that we do not introduce any new fermions except those in the
standard model. We just group them into the representations of the global
symmetry ${\bf G}$. We also see that all the fermion $SU(2)_L$ doublets
(including quarks and leptons)  are grouped into
$SU(3)$
triplets, while the $SU(2)_L$ singlets (right-handed fermions) are grouped into
$SU(3)$ anti-triplets.
 For
definiteness, we write here the explicit expressions for the Higgs fields:
\ben
\eta_{i}=\left(\bea{c}
 \eta^+_{i}\\
	            \eta^0_{i}\eea\right),\hskip.7in
\eta^{\dagger i}=\left(
 \eta^{-i}\hskip.3in
	            \eta^{0*i}\right).
\een
\ben
\Phi^{ij}=\left(\bea{c}
 \phi^{+ij}\\
	            \phi^{0ij}\eea\right), \hskip.5in
\Phi_{ij}^\dagger=\left(
 \phi^{-}_{ij}\hskip.3in
	            \phi^{0*}_{ij}\right);
\een

Now comes the Higgs potential, which we assume is  ${\bf G}$
symmetric when small
 explicit global symmetry breaking terms are neglected.  We separate the
symmetric
potential into three parts, two self-interaction parts and one cross
interaction part:
\beq
V_1(\Phi)=\delta Tr(\Phi^\dagger\Phi)+\beta(Tr(\Phi^\dagger\Phi))^2
+\alpha Tr(\Phi^\dagger\Phi)^2, \hskip.2in (\alpha+\beta > 0, \alpha+3\beta>0)
\eeq
\beq
V_2(\eta)=\mu \eta^\dagger\eta+\lambda(\eta^\dagger\eta)^2, \hskip.5in (\lambda
>0)
\eeq
\beq
V_{12}(\Phi,\eta)=\alpha_1 \Sigma_{ijk}\eta^{\dagger i}
\Phi^{jk}\Phi^\dagger_{ij}\eta_k+\alpha_2 \Sigma_{ijk}\eta^{\dagger i}\eta_k
\Phi^\dagger_{ij}\Phi^{jk}+\alpha_3\eta^\dagger\eta Tr(\Phi^\dagger\Phi).
\eeq
 The positive definite conditions for  $V_1$ and $V_2$
respectively are in parentheses. The conditions for the whole potential to be
positive definite
are more involved with the magnitudes of $\alpha_1$, $\alpha_2$ and $\alpha_3$,
compared with that of $\alpha$, $\beta$ and $\lambda$. We do not go into a
detailed discussion of them here because they are not essential to the main
subject of this note.

If both $\delta$ and $\mu$  in Eqs(4, 5) are negative, then both $\Phi$
and $\eta$ may develop VEVs. This is more so if $\alpha_3 < 0$. Let us suppose
 that one of the diagonal elements of the sixtet Higgs develops VEV[9].
 We call this component the 3-3 component,
 \beq
{\sqrt{2}}\langle \Phi^0_{33}\rangle={\it v}\sim 230 GeV.
\eeq
The effect of this VEV is to break the electroweak gauge symmetry
$SU(2)_L \times U(1)_Y$ down to $U(1)_{em}$, and to break the global
symmetry ${\bf G}$ down to $SU(2)\times U(1)^\prime $.
There are two possible patterns for $\eta$ to develop a VEV:
\beq
{\sqrt{2}}\langle \eta^0_1\rangle={\it v}^\prime,
\eeq
or\footnote{$
\langle \eta^0_2\rangle={\it v}^\prime$ is equivalent to
$\langle \eta^0_1\rangle={\it v}^\prime$. It is just a matter
of exchanging the definition of the first and second families.}

\beq
{\sqrt{2}}\langle \eta^0_3\rangle={\it v}^\prime,
\eeq
where ${\it v}^\prime$ could be complex\footnote{The condition to decide the
phase of ${\it v}^\prime$ should be discussed elsewhere.} and we assume that
the magnitude of ${\it v}^\prime$ is  smaller than ${\it v}$ (e.g. 2.5 times
smaller). When $\alpha_1+\alpha_2>0$, the first possibility is more
favorable. This VEV
pattern breaks the global symmetry further down to $U(1)$.   $U(1)_{em}$
is untouched by this VEV
because this component in Eq(8) has the same $SU(2)_L\times U(1)_Y$ property
as the component in Eq(7).

Now we are ready to discuss the Yukawa sector.
The Yukawa sector is not completely  symmetric under the global
transformations. Different Yukawa terms have different symmetry properties.
The only term that is completely symmetric is
\beq
{\it\bf L}^Y_0=\Sigma_{ij}G_0\bar\psi_L^ii\tau_2 \Phi^\dagger_{ij}U_R^j +h.c.
\eeq
which after SSB will serve the top quark a large mass, if the magnitude of
$G_0$ is close to 1. A good reason for such a large Yukawa coupling is that
this term is completely global symmetric.
The following terms are symmetric under different subgroups  of the global
symmetry ${\bf G}$
\ben
\bea{ccccc}
{\it\bf L}^Y_1&=&\Sigma_{ijk}G_1\bar\psi_L^i\eta^{\dagger j}U_R^k&
\varepsilon_{ijk}& +h.c.\\
&&&\\
{\it\bf L}^Y_2&=&\Sigma_{\alpha\beta\gamma\lambda=2,3}G_2\bar\psi_L^\alpha
\tau_{\alpha\gamma}\Phi^{\gamma\lambda}\tau_{\lambda\beta}D_R^{\beta}& &+h.c.\\
&&&\\
{\it\bf L}^Y_3&=&\Sigma_{ijk}G_3\bar\psi_L^i\eta_{j}D_R^k&\varepsilon_{ijk}&+
h.c.\\

\eea
\een
where the $G_1$ term is $SU(3) $ symmetric,  the $G_2$ term is
 $SU(2)\times U(1)^\prime\times Z_2$ symmetric, and the $G_3$ term,
 $SO(3)$ symmetric.
 Therefore, according to the principle of naturalness, these coupling
 constants
are sequentially smaller. The ratios of magnitudes of these couplings are about
\beq
  G_0:G_1\sim 5, \hskip .2in G_1:G_2\sim 7, \hskip.2in G_2:G_3\sim 2.5 .
\eeq
Note that these ratios are all of the order of $10^0$.
 The following question can be
answered here. In Eq(7) we assume that the component 3-3 of the
sixtet develops VEV. Here we assume that $SU(2)\times U(1)^\prime\times Z_2$
 symmetry of the $G_2$ term is on the 2 and 3 bases. Are these two
 assumptions
consistent ? In other
words, in the summation of the $G_2$ term, the indices $\alpha$ etc.
 run over
 two values instead of three, in order to give the smaller symmetry. Why
 must these
two values include 3 as one of them ?  Actually this is not an artificial
choice. On the
contrary, this is the only consistent choice. When the $G_2$ term
exists, there
will be an induced term (and other terms) in the potential which
is proportional to
$-|G_2|^2\Sigma_{\alpha\beta}\Phi^{\alpha \beta}\Phi^\dagger_{\alpha\beta}$
where $\alpha$ and
$\beta$ can only take 2 and 3. This induced term will make the component
$\Phi^{22}$ or $\Phi^{33}$ more favorable to develop VEV.

  The leptonic part of the
Yukawa sector has only the $G_1$ like and $G_3$ like terms, if there are not
right-handed neutrinos
\ben
\bea{cccc}
{\bf L}^Y_{lep}&=&G^l_2\Sigma_{\alpha\beta \gamma\lambda}\bar
L^\alpha\tau_{\alpha\gamma}
\Phi_{\gamma\lambda}\tau_{\lambda\beta}\,l_R^\beta&\\
&&&\\
&+&G^l_3\Sigma_{ijk}\bar L^i\eta_{j}\,l_R^k &\varepsilon_{ijk} + h.c..
\eea
\een
We will not discuss the leptonic part further.

It is easy to read off the mass matrix for the up-type quarks, which is
\ben
M^U=\left(\bea{ccc}
0&0&0\\
	             0&0&G_1{\it v}^{\prime *}\\
	              0&-G^*_1{\it v}^\prime &G_0{\it v}\eea\right).
\een
That for down-type quarks is
\ben
M^D=\left(\bea{ccc}
0&0&0\\
	              0&0&G_3{\it v}^{\prime *}\\
	              0&-G^*_3{\it v}^\prime &G_2{\it v}\eea\right).
\een
The mechanism of producing the other matrix elements is still mysterious.
In any case, these elements will be much smaller than those non-zero elements
in the corresponding mass matrices. Therefore Eqs(14, 15) are very good
approximations for
 the mass matrices.
Note that in both matrices, the 3-3 elements obtain their contributions
from the
same big VEV, which naturally explains why both are the largest matrix
elements
in their respective mass matrices.

When  diagonalizing the mass matrices we find that $V_{cb}$ can be  expressed
as ($V_{ub}=V_{td}=0$ in this stage of approximation)
\beq
V_{cb}=x-x^\prime,
\eeq
where $x$ and $x^\prime$ are two complex numbers with their values related to
the quark masses
\beq
|x|=\sqrt{m_c/m_t},\hskip.6in |x^\prime|=\sqrt{m_s/m_b}.
\eeq
Note that the $m_c$ obtained from $M^U$ is at the order of
$|(G_2{\it v}^\prime)^2/G_0{\it v}|$,
which is much smaller than the smaller elements in $M^U$. Similarly,
 $m_s$ is much smaller than the smaller elements in $ M^D$. Therefore,
small corrections to the zeroes in the mass matrices (14, 15) may cause the
mass
formulas for $m_c$ and $m_s$ to change an appreciable fraction, and
consequently to change the
formulas for $x$ and $x^\prime$ appreciably, which may or may not improve the
value of $V_{cb}$ in Eqs(16, 17).

The masses of the psuedo-goldstone bosons in this model come mainly from the
explicit
 symmetry breaking terms in the Higgs potential in this model.  These
 asymmetric
  terms are supposed to break
the continuous global symmetry completely, in order to avoid any goldstone
particles that do not obtain masses after SSB. There are many possible terms,
for example
$\mu^\prime|\Sigma_i \eta_i|^2$. Once the condition $\mu^\prime <<\mu$ and
$ \delta$
is satisfied, the basic picture discussed above will not be disturbed very
much.
The pseudo-goldstones obtain their masses also from the Yukawa interaction
terms which break the corresponding global symmetry. However this contribution
is much smaller than that from the explicit symmetry breaking terms in the
Higgs potential.
A limit for the masses of $PGBs$ which mediate the $b\rightarrow s$ transition
should in principle be able to be obtained from the data on $B_s-\bar B_s$
mixing and
$b\rightarrow s+\gamma$. The existence of flavor-changed neutral currents at
the tree level are unavoidable because both up-type and down-type quarks obtain
their masses from two different Higgs multiplets. However, a
further discussion of this is out of the
scope of this note.

Let us instead discuss briefly the interesting discovery channels of the
$PGBs$.
There are eight neutral $PGBs$\footnote{There is no charged $PGB$  because
the
global symmetry for the charged particles does not spontaneously break, for
only
  the neutral components develop VEVs.}, when the global symmetry
 $SU(3)\times U(1)$
spontaneously breaks down by the VEVÕs in Eqs(7, 8) to $U(1)$.
As pointed out at the beginning of this note, $PGBs$ cannot be discovered by
$Z \rightarrow Z^*+PGB$ or  $Z^* \rightarrow Z+PGB$ because there is no such
 coupling. The branching
ratio of double $PGB$ production on the $Z$ peak is too tiny because it is a
third order effect and the phase space for the final state is too small. The
most interesting channel is top decay[1]
and we have approximately
\beq
\frac{\Gamma(t\rightarrow c+PGB)}{\Gamma(t\rightarrow b +W^+)}\sim |x|^2(
\frac{m_t}{m_W})^2,
\eeq
if the relevant PGB is appreciably lighter than the top quark.
So the decay into a pseudo-goldstone boson may be an appreciable channel of top
decay. Such produced  $PGB$
will then decay into $b+s $ or $c+u$ etc. For $PGBs$ which do not mediate
flavor-changed neutral currents, their coupling constants to
the light quarks are  larger than that in the minimal standard model, because
of
the smallness of ${\it v}^\prime$. Therefore they have a better
chance to be found in hadron colliders.

In conclusion, we have presented here an alternative model which relates the
scalar spectrum with the masses of quarks and leptons. $SU(3) \times U(1)$ is
used as the global horizontal symmetry. Left-handed fermions are in the triplet
 representations while right-handed fermions are in the anti-triplet
representations. Both the Higgs triplet and sixtet are introduced to provide
VEVs with different values. The complete global symmetry is not secured in
the Yukawa sector. $SU(3)\times U(1)$ and $SU(2)\times U(1)^\prime\times Z_2$
symmetric
Yukawa couplings are
allowed for the sixtet Higgs and $SU(3)$ and $SO(3)$
symmetric Yukawa couplings are allowed for the triplet Higgs. In this way we
obtain second order mass matrices for both up- and down-type quarks.

The author thanks Dr. Chung Kao, and Dr. Zhi-Zhong Xing for discussions. He
also acknowledges that Dr. Steven Weinberg has read the manuscript.

This work is in part supported by  the Department of Energy of the United
States under contract number DE-FG03-95 ER40914/A00.

\clearpage
\begin{center}
{\bf Table 1}
\end{center}
\vskip.15in
$$\bea{ccccc}
 \hskip.5in object\hskip.5in &\hskip.2in SU(3) \hskip.2in &\hskip.2in
SU(2)_L\hskip.2in &\hskip.2in U(1)^*\hskip.2in &\hskip.2in
U(1)_Y\hskip.2in\\
&&&&\\

\psi_{Li}&3&2&1&\frac{1}{6}\cr
&&&&\\

L_i&3&2&\frac{2}{3}&-\frac{1}{2}\cr
&&&&\\

U_R^i&
\bar 3 &1&-1&\frac{2}{3}\cr
&&&&\\

D_R^i&\bar 3 &1&-1&-\frac{1}{3}\cr
&&&&\\

l_R^i&\bar 3&1& -\frac{4}{3}&-1\cr
&&&&\\

\Phi^{ij}& \bar 6&2& -2&\frac{1}{2}\cr
&&&&\\

\eta_{i}&  3&2& 1&\frac{1}{2}\cr

\eea
$$

\hskip.1in {\small $^*$~ The $U(1)$ charge $\xi=I-L/3$, where I is the
representation index of $SU(3)$ and \\
{}~~~~ $L$ is
the lepton number.}
\clearpage
{\bf references}
\begin{itemize}
\begin{enumerate}
\item The CDF Collaboration, FERMILAB-pub-95/022-E;
S. Abachi, et al, The D0 Collaboration, Pub-95/028-E.

\item S. Weinberg, Phys. Lett. 19, 1264(1967);  A. Salam, in {\it Elementary
Particle Theory,} ed. N. Svartholm (Almquist and Wilksells, Stockholm, 1969)
 p367; S. L. Glashow, Nucl. Phys. 22, 579(1961).

\item S.L. Glashow, J. Iliopoulos, and L. Maiani, Phys. Rev. D2, 1285(1970).
\item N. Cabibbo, Phys. Rev. Lett. 10, 531(1963); M. Kobayashi, and T. Maskawa,
Prog. Theor. Phys. 49, 652(1973).

\item For the bounds on the Higgs masses in the minimal standard model and
some frequently discussed models see, the Particle Data Group, Phys. Rev.
D50(1994)No.3-I.
\item S. Weinberg, UTTG-05-91, (Contribution to a volume in honor of Baqi Beg).
\item H. Fritzsch, Phys. Lett. 73B, 317(1977).
\item For example, see R. M. Xu, Phys. Rev. D44, 590(1991); N.G. Deshpand
and X. G. He, Phys. Rev. D49, 4812(1994).
\item D. D. Wu, Nucl. Phys. 199B, 523(1981).
\end{enumerate}
\end{itemize}
\end{document}